\documentclass{moriond}

\usepackage{amsmath}
\usepackage{amssymb}

\def\Journal#1#2#3#4{{#1} {\bf #2}, #3 (#4)}

\def\PRL{\em Phys. Rev. Lett.}
\def\PRD{{\em Phys. Rev.} D}

\def\APJ{{\em Astrophys. J.}}
\def\AJ{{\em Astron. J.}}
\def\AA{{\em Astron. Astrophys.}}

\def\lcdm{\Lambda\rm CDM}

\def\s8{S_8}

\newcommand{\Photo}{}

\begin{document}
\vspace*{4cm}
\title{COSMOLOGY SINCE THE FIRST ASTRO/COSMO MORIOND MEETING:\\
THE EMERGENCE OF THE BIG BANG 2.0}

\author{A. Blanchard}

\address{Institut de Recherche en Astrophysique et Plan\'etologie (IRAP),\\
Universit\'e de Toulouse, France}

\maketitle\abstracts{
This paper presents a necessarily incomplete review of the evolution of cosmology since the first Astro/Cosmo Moriond meeting in 1981. I trace the journey from the classical Big Bang model based on three pillars---universe expansion, primordial nucleosynthesis, and the cosmic microwave background---to the modern $\Lambda$CDM paradigm and the discovery of cosmic acceleration. I discuss major observational milestones: the COBE discovery of CMB fluctuations, the CMB measurements of the flat universe, the pivotal discovery of accelerated expansion through Type Ia supernovae and the emergence of precision cosmology with Planck. I  review current tensions in cosmological parameters, particularly the Hubble tension and $\s8$ discrepancies, and discuss future prospects from large-scale structure surveys like DESI. The emergence of ``Big Bang 2.0'' reflects the profound paradigm shift from a model based on standard physics to a dynamical cosmos dominated by dark matter and dark energy, the description of which requires a physics that has yet to be developed and validated.}

\section{Introduction: Forty-Five Years of Transformation}

Cosmology has undergone a remarkable transformation over the past 45 years. Since the first Astro/Cosmo Moriond meeting in 1981, convened by Jean Audouze and Jean Tr\^an Thanh Van, the field has evolved from a data-poor to a data-rich enterprise, delivering unprecedented precision measurements of cosmic parameters with accuracy reaching the sub-percent level. This review traces the major milestones and paradigm shifts that have defined this evolution, from the establishment of the classical Big Bang to the discovery of cosmic acceleration and the consolidation of a new vision
which requires a new physics leading to the $\Lambda$CDM model, but which is now facing a number of "tensions".

The Big Bang model, first conceived by Georges Lema\^itre, is grounded in General Relativity and standard physics. It rests on three foundational pillars: (1) expansion of the universe, described by the Hubble-Lema\^itre relation $V = H_0 D$ ($V$ is inferred from the shift in wavelength, the redshift $z$ given by  $1+z = \lambda_0/\lambda$); (2) primordial nucleosynthesis, the synthesis of light elements ($^4$He, D, $^3$He, $^7$Li) in the first few minutes after the Big Bang; and (3) the cosmic microwave background, discovered in 1965 by Penzias and Wilson as a thermal radiation with temperature $T \approx 2.73$ K. By around 1970, this picture was largely established and provided a coherent framework for understanding cosmic evolution from fractions of a second after the Big Bang to the present day, some 13.8 billion years later.

However, fundamental questions remained unanswered: What is the origin of large-scale structure in the universe---the galaxies, clusters, and filaments we observe today? Why is there an asymmetry between matter and antimatter, given that the Big Bang should have produced equal amounts of each? What comprises the dark matter that appears to dominate the gravitational dynamics of galaxies and clusters? How can we explain the absence of the colossal amount of energy in the quantum vacuum predicted by quantum field theory?

The 45 years since the 1981 Moriond meeting have provided remarkable answers to many of these questions, while raising new and even more profound puzzles. The discovery of cosmic acceleration, the precise measurement of cosmic geometry, and the revelation of a universe dominated by mysterious dark matter and dark energy have fundamentally reshaped our understanding of nature.

\section{Paradigm Shifts: Inflation and Cold Dark Matter (1981--1995)}

\subsection{The Inflationary Paradigm}

The 1981 proposal by Allan Guth of cosmic inflation represented a paradigm shift of fundamental importance~\cite{guth1981}. Guth recognized that the classical Big Bang model faced two critical problems: the horizon problem (why is the universe so uniform when different regions could not have been in causal contact?) and the flatness problem (why is the spatial curvature so close to zero, requiring an extraordinary degree of fine-tuning?). Inflation---an epoch of extraordinarily rapid exponential expansion in the very early universe---provides elegant solutions to both problems.

During inflation, the universe expands by many orders of magnitude in a fraction of a second, driven by the potential energy of a scalar field (the inflaton). This expansion stretches quantum fluctuations to macroscopic scales, providing the seed density perturbations that later grow through gravitational instability to form galaxies and cosmic structure. Inflation occurs at extraordinarily high energy scales, possibly $\sim 10^{15}$ GeV, just below the Planck scale where quantum gravity effects become important.

\subsection{Cold Dark Matter and Structure Formation}

Jim Peebles~\cite{peebles1982}, building on earlier work with J.T. Yu~\cite{peebles1970}, examined the observational consequences of cold dark matter (CDM) in a cosmological context. Peebles developed and refined Boltzmann codes that track the evolution of density perturbations in the early universe, accounting for the interactions between photons, baryons, and dark matter. These codes predict both the temperature power spectrum of the CMB and the matter power spectrum governing large-scale structure.

The Cold Dark Matter picture was compelling: non-baryonic dark matter dominates the gravitational evolution of the universe, while baryons (protons, neutrons, and their bound states) constitute only $\sim 5\%$ of the total matter density. The initial density perturbations, presumably generated during inflation, have a scale-invariant spectrum characterized by the spectral index $n_s \approx 1$.

\subsection{The first "large" redshift survey}

The first redshift survey conducted to study large-scale structures in the universe was the CfA survey \cite{TonryDavis1979}. Its release triggered numerous investigations into the statistical nature of the galaxy distribution as well as  dynamics on large scale \cite{DavisPeebles1983}.

\subsection{The COBE Discovery (1992)}

COBE was the first satellite dedicated to the observations of the Cosmic Microwave Background. The COBE observations  established the blackbody nature of the CMB spectrum to exquisite precision, confirming the thermal history of the Big Bang. This was the crowning moment for classical Big Bang picture. 

The 1992 discovery by the COBE DMR instrument of temperature fluctuations in the CMB at the level of $\Delta T / T \approx 10^{-5}$ was revolutionary. These tiny temperature variations, with amplitude of only microkelvin, provide direct evidence that quantum fluctuations in the very early universe have been stretched to cosmological scales by inflation. Stephen Hawking famously called this ``the most significant discovery of the century, if not of all time.'' The small departures from isotropy revealed by COBE set the stage for more detailed investigations of CMB anisotropies at smaller angular scales.

\subsection{Challenges and the $\Lambda$CDM Breakthrough (1992--1995)}

By the early 1990s, observations began to constrain this picture in surprising ways. The APM (Automatic Plate Measurement) survey measured the angular correlation function of galaxies, providing information about the clustering of matter in the nearby universe \cite{APM} inconsistent with the shape expected in CDM model with $\Omega_m = 1$. A bit latter,  measurements of the baryon mass fraction in clusters of galaxies obtained from x-ray observations, lead to a ratio of baryonic to total mass around  $\sim 15\%$ in clusters. If this ratio is universal (a reasonable assumption), it implies that the total matter density of the universe is only $\Omega_m \approx 0.3$ \cite{White93},
 and not $\Omega_m = 1$, as was previously thought to resolve the flatness problem within inflationary theory.
 
 Although Peebles pointed out quite early that inflation implies $\Omega_m + \Omega_\Lambda= 1$ \cite{Peebles1984}, it is only after the above facts that 
a few cosmologists  began to consider the possible existence of a non-zero cosmological constant\cite{Efstathiou1990}. Ostriker and Steinhardt (1995)~\cite{ostriker1995} proposed to solve the  above apparent contradiction by  proposing the $\Lambda$CDM model: a spatially flat universe ($\Omega_{tot} = 1$) with matter density $\Omega_m \approx 0.3$ and a cosmological constant or ``dark energy'' component contributing $\Omega_\Lambda \approx 0.7$. The cosmological constant, representing the energy density of the vacuum, provides the missing energy density needed to flatten the geometry while keeping the matter density consistent with observations. This model was initially speculative, but it would prove remarkably prescient.

\subsection{Evidence for Flatness: The CMB Acoustic Peak (1996--2000)}

Pivotal evidence for a flat universe emerged from ground-based CMB experiments in 1996 \cite{Charley1996}. Several experiments detected a pronounced feature in the CMB power spectrum---an acoustic peak arising from sound waves in the early plasma. The location of this peak in angular scale depends sensitively on the geometry of the universe: in a flat universe, the peak appears at a characteristic angular scale determined by the sound horizon at recombination, while in a curved universe, projection effects shift the peak location.

The 1996 observations, comforted with more precise measurements from the Boomerang (2000) and Archeops (2003) experiments, provided direct evidence that the universe is indeed spatially flat, supporting the $\Lambda$CDM model. Subsequently, the higher-precision measurements from WMAP (2003--2010) and Planck (2009--2018) confirmed flatness to exquisite precision, constraining the spatial curvature to $\Omega_K = 0.001 \pm 0.002$, fully consistent with zero.

\section{The Discovery of Cosmic Acceleration: 1998 and Beyond}

\subsection{Type Ia Supernovae and the Hubble Diagram}

The measurement of Type Ia supernovae (SNIa) at cosmological distances provides a powerful tool for measuring cosmic distances and the expansion history of the universe. These events are understood as thermonuclear explosions of white dwarf stars in binary systems, where material from a companion star accretes onto the white dwarf until thermonuclear runaway occurs. The peak brightness of SNIa explosions is approximately standardizable, making them ``standard candles'' for cosmological distance measurements.

Two independent collaborations---the Supernova Cosmology Project (led by Saul Perlmutter) and the High-$z$ Supernova Search Team (led by Adam Riess)---conducted surveys to discover and measure SNIa at high redshifts ($z = 0.3$--$1.0$). By 1998, each collaboration had accumulated $\sim 50$ SNIa with precise light curves and redshift measurements. When they plotted these data on the Hubble diagram (luminosity distance versus redshift), they made a startling discovery: the high-redshift supernovae were significantly dimmer than expected.

In a universe dominated by matter, the expansion would be decelerating due to gravity. The Hubble diagram of  supernovae at high redshift can be easily evaluated in cosmological models. The observed supernovae were fainter than expected in any decelerated universe. The data were consistent, instead, with a universe that is accelerating. Although this effect might have been due to an astrophysical phenomenon (such as an evolutionary process), it came at just the right moment to provide the missing piece of the puzzle.

This discovery, published in 1998 by both collaborations~\cite{perlmutter1999,riess1998}, was revolutionary. It implied that the universe contains a component with negative pressure---a repulsive ``force'' that causes the expansion to speed up rather than slow down. This component, termed ``dark energy,'' constitutes $\sim 70\%$ of the energy density of the present-day universe.

\subsection{The Nature of Dark Energy: The Cosmological Constant and Beyond}

The simplest explanation for cosmic acceleration is Einstein's cosmological constant ($\Lambda$), originally introduced in 1919 to permit a static universe and later discarded as unnecessary when Hubble's observations revealed cosmic expansion. Lema\^itre (1934) provided a modern reinterpretation: the cosmological constant can be understood as the energy density of the vacuum, a fundamental prediction of quantum field theory.

However, the theoretical prediction for vacuum energy density is catastrophically wrong. Estimations based on quantum field theory suggest the vacuum energy should be $\sim 10^{113}$ joules per cubic meter, whereas observations indicate $\sim 10^{-9}$ joules per cubic meter. This discrepancy of $\sim 10^{120}$ is the ``cosmological constant problem,'' arguably the most serious unresolved problem in theoretical physics.

Alternative explanations for cosmic acceleration include:

{\it Dynamical dark energy} (quintessence): Rather than a constant, the equation of state parameter $w = P/\rho$ might evolve with redshift. Parameterizations such as the Chevallier-Polarski-Linder (CPL) form, $w(a) = w_0 + w_a(1-a)$, allow for evolving dark energy while remaining observationally tractable.

{\it Modified gravity}: Modifications to General Relativity at cosmological scales could mimic cosmic acceleration without invoking dark energy. Examples include $f(R)$ gravity, scalar-tensor theories, and other alternatives. However, most modified gravity theories face observational constraints from large-scale structure and other cosmological tests.


\section{Precision Cosmology: CMB and Large-Scale Structure}

\subsection{The CMB as a Precision Probe of Early Universe Physics}

The cosmic microwave background provides a snapshot of the universe at recombination, when the universe had cooled to $T \approx 3000$ K and became transparent to radiation. This corresponds to a redshift $z \approx 1100$, roughly 380,000 years after the Big Bang. Photons decoupled from matter at this epoch and have propagated freely to us, 
thus providing a picture of the universe at that early time.

Modern CMB experiments (Planck, WMAP, ACT, SPT, SO) have measured temperature anisotropies to exquisite precision. The Planck satellite, operating from 2009 to 2013, achieved high sensitivity  across a wide range of angular scales. Analysis of the temperature power spectrum $C_\ell$ yields precise constraints on six fundamental cosmological parameters of the $\Lambda$CDM to within a few per cent:

\begin{itemize}
\item Baryon density: $\Omega_b h^2 = 0.0224 \pm 0.0001$
\item Cold dark matter density: $\Omega_c h^2 = 0.1201 \pm 0.0012$
\item Hubble constant: $H_0 = 67.4 \pm 0.5$ km/s/Mpc
\item Age of the universe: $t_0 = 13.813 \pm 0.026$ Gyr
\item Scalar spectral index: $n_s = 0.9649 \pm 0.0042$
\item Optical depth to reionization: $\tau = 0.0544 \pm 0.0073$
\end{itemize}


\subsection{Large-Scale Structure: Baryon Acoustic Oscillations and Galaxy Clustering}

Complementary constraints on cosmological parameters come from the distribution of galaxies and other tracers of matter in the late-time universe. Large spectroscopic surveys (SDSS, 2dF, DESI) have measured redshifts for millions of galaxies, enabling detailed maps of cosmic structure and precise measurements of galaxy clustering.

A particularly powerful observational signature is the baryon acoustic oscillation (BAO), a characteristic scale imprinted by sound waves propagating in the primordial plasma. These sound waves reflect the balance between pressure forces and gravitational attraction; they freeze in at recombination, leaving an imprint in the CMB as well as on the clustering of matter that persists to the present day. This feature, the BAO scale, provides a ``standard ruler'' of known size, enabling precision measurements of cosmic distances and the expansion rate at different redshifts.

The SDSS data revealed this BAO feature in the correlation function of galaxies at a scale of $\sim 100$ Mpc/h, in excellent agreement with predictions from $\lcdm$ adjusted to Planck CMB measurements. More recently DESI has extended BAO measurements to higher accuracy, tracing the expansion history and testing consistency between early- and late-universe probes.

It is a truly remarkable achievement of the $\lcdm$ model that it can account for the universe when it was 0.002\% of its current age (thanks to the CMB) and is in complete agreement with everything we know about the structure of the universe today (and its recent past) more than 13 billion years later, even though the universe has expanded by a factor of around 1,100. 

\section{Current Tensions and Hints of New Physics}

Tensions now appear because some of the parameters determined from the CMB differ from their values inferred from the local universe by a few \%. This illustrates the era of "precision cosmology", a major shift from the time of Moriond 1981. 

\subsection{The Hubble Tension}

One of the most significant puzzles in modern cosmology is a discrepancy between measurements of the Hubble constant $H_0$ from different methods. The local distance ladder, based on Cepheid variable stars and Type Ia supernovae, yields $H_0 = 73 \pm 1$ km/s/Mpc (SH0ES collaboration). In contrast, fitting the $\Lambda$CDM model to Planck CMB data and BAO measurements yields $H_0 = 68.17 \pm 0.28$ km/s/Mpc \cite{DESI2025}. This $\sim 5\sigma$ discrepancy is highly significant statistically.

Several possibilities have been proposed to explain this tension,  including
 Early dark energy (EDE), which could increase the expansion rate in the early universe, altering the sound horizon scale and lowering the inferred $H_0$ from CMB data, bringing it into agreement with local measurements. However, no convincing solution has yet emerged, and a possibility that remains is  unaccounted systematics ("unknown unknown") in the distance ladder based on Cepheids.

Resolution of this tension is one of the highest priorities in observational cosmology.

\subsection{The $\s8$ Tension and Matter Power Spectrum}

Weak gravitational lensing and measurements of galaxy cluster abundances constrain the amplitude of matter fluctuations on scales of $\sim 8$ $h^{-1}$Mpc, parameterized as $\sigma_8$ or equivalently through $\s8 = \sigma_8 (\Omega_m/0.3)^{0.5}$. Observations from DES, KiDS, and other weak lensing surveys yield $\s8$ values somewhat lower ($\sim 2\sigma$) than predicted by Planck CMB measurements. While this tension is less severe than the $H_0$ tension and may be weakening with more recent data and improved systematic error accounting, it suggests possible tension in the growth of structure.

\section{Future Prospects: Next-Generation Surveys and Experiments}

\subsection{Current and future facilities}

Several ambitious projects already active or  in development will transform cosmology over the next decade.

{\it DESI} has published results based on the first three years of observations. The initial survey originally planned to run for five years, will continue until at least 2028.

{\it Euclid} (ESA, launch 2023) will conduct a weak gravitational lensing survey of $\sim 2$ billion galaxies, measuring the shapes of galaxy images distorted by the gravitational lensing effect of intervening dark matter. The combination  of galaxy clustering in the photometric survey and their cross-correlation, the 3x2pt analysis,  potentially provides exquisite constraints. In addition Euclid  will collect a large spectroscopic redshifts survey,  enabling additional dark energy constraints through measurements of cosmic distances and the growth of structure.

{\it Roman Space Telescope} (NASA) will extend supernova cosmology to higher redshifts ($z \approx 2$), providing independent measurements of cosmic acceleration and dark energy evolution beyond the range accessible to ground-based surveys.

{\it LSST} (Vera C. Rubin Observatory, 2025--) will conduct for ten years a deep, wide-field imaging survey of the entire sky, discovering millions of supernovae, measuring weak lensing on unprecedented scales, and enabling transient astronomy on new frontiers.

{\it  Simons Observatory} (SO) as well as the LiteBird satellite will achieve unprecedented sensitivity to CMB temperature and polarization, enabling precision measurements of inflation through detection of primordial gravitational waves via B-mode polarization, if these waves are energetic enough to be observable. Detection would probe the inflationary energy scale, potentially revealing physics at $10^{15}$--$10^{16}$ GeV---extraordinarily high energies never accessible to terrestrial experiments.

\section{Recent Observational Results and Cosmic Parameter Constraints}

\subsection{DESI Results (2025)}

The Dark Energy Spectroscopic Instrument (DESI) completed its first year of observations in 2025, measuring redshifts for nearly 5 million galaxies and providing unprecedented constraints on large-scale structure and dark energy. DESI combines photometric imaging from the Dark Energy Camera with a fiber spectrograph capable of measuring galaxy spectra simultaneously for thousands of objects.

DESI's early results reveal:

{\it BAO measurements at multiple redshifts} provide independent measurements of cosmic distances and the expansion rate history. Measurements of the BAO scale at $z \approx 0.5$, $z \approx 1$, and higher redshifts yield the expansion rate $H(z)$ at different epochs, enabling tests of dark energy models.

{\it Dark energy equation of state constraints}: the combination of BAO, growth rate, and weak lensing measurements constrain the effective equation of state parameter $w$. DESI results show hints of evolving dark energy, with $w(z) \neq -1$ tentatively suggested at the $\sim 2-4\sigma$ level--- but with no hint to  solve cosmological tensions.

{\it Tension implications}: The DESI results, while consistent with $\Lambda$CDM within current uncertainties, continue to hint at tensions with early- universe physics, as inferred from Planck and DESI,  and the late time universe with SH0ES measurements. It is crucial to identify whether these tensions represent genuine deviations from $\Lambda$CDM or result from unrecognized systematics.

\subsection{Weak Lensing from DES and KiDS}

Weak gravitational lensing---the subtle bending of light from distant galaxies by the gravitational field of intervening dark matter---provides a direct probe of the dark matter distribution. The Dark Energy Survey (DES)\cite{DES6yr} and the Kilo-Degree Survey (KiDS)\cite{KiDSL} have measured the shapes and positions of hundreds of millions of galaxies, enabling tomographic maps of dark matter at multiple redshifts.

The $\s8$ tension seems to be weakening with more recent data and improved systematic error accounting. DES weak lensing measurements yield constraints on $\s8 = 0.789\pm 0.012$ that are lower than CMB predictions at the $\sim 2\sigma$ level while KiDS concluded to $\s8 = 0.814\pm 0.012$. This tension, if real,  could indicate: (1) new physics affecting the growth of structure, (2) unaccounted systematics in intrinsic alignment or photometric redshifts, or (3) statistical fluctuations in the current datasets.

\subsection{Cross-Correlations Between Probes}

Modern cosmology increasingly relies on cross-correlations between different observational probes. The integrated Sachs-Wolfe (ISW) effect can be measured by cross-correlating the CMB with large-scale structure maps, yielding independent constraints on dark energy at recent epochs ($z <~ 2$). These measurements confirm the consistency of dark energy properties across different cosmic eras and provide tests of modified gravity.

Cross-correlations between weak lensing maps from imaging surveys and galaxy clustering from spectroscopic surveys yield powerful constraints on the relationship between dark matter and galaxies: the cross correlation between the two field gives all its power to the 3x2pt method \cite{Euclid3x2ptf}.

\section{Paths to Resolution: Early Dark Energy and Modified Gravity}

\subsection{Early Dark Energy Solutions}

Several models have been proposed in which a component of dark energy is significant in the early universe (at $z > 1000$), where it temporarily increases the expansion rate, early dark energy \cite{EDE}. This increased early expansion changes the sound horizon scale inferred from the CMB, increasing the predicted $H_0$ value and potentially bringing it into agreement with local measurements.

However, early dark energy models face significant constraints. To solve the $H_0$ tension, they require the dark energy density parameter at early times to reach $\Omega_{e} \sim 0.05$---a substantial fraction of the total energy density---for a limited window around $z \sim 3000$. Constraints from primordial nucleosynthesis, the CMB power spectrum, and large-scale structure limit the available parameter space. Nevertheless, early dark energy remains a promising avenue for resolving current tensions.

\subsection{Modified Gravity Theories}

Alternative approaches propose modifications to General Relativity at cosmological scales. These theories alter the relationship between matter distribution and spacetime curvature, potentially affecting the expansion history and the growth of structure in ways that could reconcile current observations.

Examples include $f(R)$ gravity theories where the Einstein-Hilbert action is modified by a function of the Ricci scalar; scalar-tensor theories (like Brans-Dicke theory) where gravity is mediated by both spacetime curvature and a scalar field; and more exotic proposals like bigravity or massive gravity.

Most modified gravity theories are now constrained by multiple observations: the cosmic microwave background power spectrum, weak lensing measurements, large-scale structure, and solar system tests (which require that deviations from General Relativity be screened at small scales). The degeneracy between modified gravity and dark energy effects can be broken through precision measurements of cosmic growth and expansion rates.

\section{Theoretical Implications and the Path to New Physics}

Despite the remarkable success of $\Lambda$CDM, several fundamental questions remain unanswered:

{{\it What is dark matter?} While weakly interacting massive particles (WIMPs) remain candidates, alternative possibilities including axions, sterile neutrinos, and primordial black holes are under investigation. Direct detection experiments, collider searches, and cosmological observations provide complementary constraints.

\it What is dark energy?} Is it the cosmological constant, dynamical dark energy, or a modification to General Relativity? The $H_0$ tension and DESI hints of evolving dark energy suggest that the simple cosmological constant may be insufficient.

{\it Can we detect primordial gravitational waves?} The measurement of the tensor-to-scalar ratio $r$ would confirm inflation and probe physics at extraordinarily high energy scales. SO and LiteBird aim for sensitivities to $r \sim 10^{-3}$.
Direct detection of the primordial gravitational wave background can also be considered.

{\it What is the origin of the matter-antimatter asymmetry?} Current measurements precisely quantify the baryon-to-photon ratio $\eta_B \approx 6 \times 10^{-10}$, but its physical origin remains mysterious. Sakharov's conditions for baryogenesis require baryon number violation, CP violation, and departure from thermal equilibrium---all potentially observable through cosmological and laboratory experiments.

{\it Are there additional relativistic species?} Measurements of the effective number of neutrinos $N_{\rm eff}$ constrain the presence of additional light particles in the early universe. Standard model predictions give $N_{\rm eff} = 3.044$; deviations would indicate new physics such as additional light neutrinos or other exotic particles.

The resolution of these open questions will certainly require new observational facilities and theoretical frameworks. The next generation of experiments promises to make significant progress on all these fronts.

\section{Systematic Challenges and the Path Forward}

As precision improves, controlling systematic uncertainties becomes paramount. Current and future surveys face several critical systematic challenges:

{\it Photometric redshift errors}: Large imaging surveys must achieve photometric redshift accuracies of $\sim 0.003(1+z)$ to avoid biasing weak lensing and clustering measurements. This requires multi-band photometric observations, photometric calibration across surveys, and validation with spectroscopic samples.

{\it Intrinsic alignment}: Galaxy shapes are not randomly oriented; they can be aligned with the surrounding matter distribution or the tidal field. This ``intrinsic alignment'' contaminates weak lensing measurements and must be modeled with exquisite precision, typically requiring detailed numerical simulations.

{\it Baryon feedback}: Non-gravitational physics (supernovae feedback, active galactic nuclei) affects the dark matter distribution on scales of $\sim 1$--$10$ Mpc. Understanding and modeling these effects is crucial for extracting cosmological information from small-scale structure and avoid bias in the determination of cosmological parameters..

{\it Calibration across surveys}: Future cosmology will combine data from multiple surveys (Planck, DESI, LSST, Euclid, Roman). Ensuring photometric calibration consistency and understanding relative systematic biases between surveys is a significant technical challenge. Cross-calibration using overlapping sky regions, overlap between different photometric systems, and validation against spectroscopic redshift samples will be essential for extracting maximal information while minimizing systematic errors that could bias derived cosmological parameters.

\section{The Role of International Collaboration and Data Management}

The Moriond cosmology meetings have served as a unique forum for the development of modern cosmology for 45 years. Starting in 1981 with a handful of participants, the meetings have grown to accommodate hundreds of researchers from around the world. The informal setting of La Thuille has fostered collaborative discussions and enabled rapid dissemination of new ideas and results.

Major breakthroughs in cosmology have been announced and debated at Moriond: the proposal of inflation and CDM as solutions to fundamental problems; the discovery of the flat universe through CMB acoustic peak measurements; the revelation of cosmic acceleration through Type Ia supernovae; and the first precision constraints from WMAP and Planck. Looking forward to 2028 and beyond, we anticipate that Moriond will continue to serve as the premier venue for discussing transformative developments in cosmology.

\subsection{The Next Generation of Cosmological Questions}

As we contemplate the next phase of cosmological research, several profound questions guide our investigations: What is the nature of dark matter, and can direct detection experiments or collider searches complement cosmological constraints? How can we definitively distinguish between a cosmological constant and dynamical dark energy? What combination of observational techniques will finally resolve the Hubble tension? Can we detect primordial gravitational waves and probe the energy scale of inflation? 

The answers to these questions will emerge from the synergistic combination of multiple observational techniques. No single probe---not CMB measurements, not supernovae, not weak lensing, not galaxy clustering, not gravitational waves---can by itself provide definitive answers. Rather, the cross-correlation and mutual validation of diverse independent measurements will illuminate the path forward.

\section*{Conclusions and Outlook}

The past 45 years since the first Astro/Cosmo Moriond meeting have witnessed a revolution in cosmological understanding. We have transitioned from qualitative models to quantitative predictions, tested to exquisite precision. The discovery of cosmic acceleration stands as one of the most profound in modern science, reshaping our understanding of the universe's fate and the fundamental laws governing gravity.

Yet mysteries remain. The tensions in cosmological parameters---particularly the Hubble tension---hint that the standard $\Lambda$CDM model may be incomplete. The next decade will be transformative, as new observations from DESI, Euclid, LSST, SO, LiteBird and other facilities probe the universe with unprecedented precision, potentially revealing new physics in dark energy, dark matter, or gravity itself.

The Moriond meetings have served as a unique and invaluable forum for discussing cutting-edge developments in cosmology. As we look forward to future meetings, we anticipate equally remarkable discoveries that will continue to reshape our understanding of the cosmos. The insights gained from resolving current cosmological tensions will almost certainly revolutionize our understanding of fundamental physics and the nature of reality itself.

\section*{Acknowledgments}

The author thanks the organizers of Moriond Cosmology 2026 for the opportunity to present this review. The author thanks  Isaac Tutusaus for his comments. The preparation of this work benefited from discussions with Claude (Anthropic) and with Euria (Infomaniak).

\end{document}